\documentclass[reqno,12pt,a4paper]{amsart}

\usepackage{mathrsfs}
\usepackage{amssymb}
\usepackage{amsfonts}
\usepackage{latexsym}
\usepackage{amsthm}
\usepackage{graphicx}
\def\lb{\label}

\newcommand{\er}[1]{\textrm{(\ref{#1})}}
\def\lb{\label}

\theoremstyle{plain}

\newtheorem{theorem}{\bf Theorem}[section]
\newtheorem{lemma}[theorem]{\bf Lemma}

\theoremstyle{remark}

\renewcommand{\a}{\alpha}           

\renewcommand{\b}{\beta}            \newcommand{\cB}{\mathcal{B}}

\newcommand{\ve}{\varepsilon}

\renewcommand{\l}{\lambda}

\newcommand{\m}{\mu}

           \newcommand{\cR}{\mathcal{R}}

\newcommand{\F}{\Phi}

\def\Z{\mathbb{Z}}
\def\R{\mathbb{R}}
\def\C{\mathbb{C}}
\def\T{\mathbb{T}}
\def\N{\mathbb{N}}

\def\qqq{\qquad}
\def\qq{\quad}

\let\ge\geqslant
\let\le\leqslant

\newcommand{\ca}{\begin{cases}}
\newcommand{\ac}{\end{cases}}
\newcommand{\ma}{\begin{pmatrix}}
\newcommand{\am}{\end{pmatrix}}

\renewcommand{\[}{\begin{equation}}
\renewcommand{\]}{\end{equation}}

\def\pa{\partial}

\def\no{\noindent}
\def\ol{\overline}
\def\iy{\infty}

\def\/{\over}
\def\ts{\times}

\def\ss{\subset}

\def\lan{\langle}                \def\ran{\rangle}

\def\Tr{\mathop{\rm Tr}\nolimits}
\def\const{\mathop{\rm const}\nolimits}

\def\BBox{\hspace{1mm}\vrule height6pt width5.5pt depth0pt \hspace{6pt}}
\def\wh{\widehat}
\def\as{\text{as}}
\def\where{\text{where}}
\def\1{1\!\!1}

\begin{document}

\title [{Trace formula for fourth order operators on the
circle}] {Trace formula for fourth order operators on the circle}

\date{\today}

\author{Andrey Badanin}
\address{Andrey Badanin,
Northern (Arctic) Federal University,
Arkhangelsk, Russia, Northern Dvina emb, 17, e-mail: an.badanin@gmail.com
}

\author{Evgeny Korotyaev}
\address{Evgeny Korotyaev,
Mathematical  Physics Department, Faculty of Physics, Ulianovskaya
2, St. Petersburg State University, St. Petersburg,  198904 Russia,
e-mail: korotyaev@gmail.com }

\maketitle

\begin{abstract}
We determine the trace formula for the fourth order operator on the circle.
This formula is similar to the famous trace formula for the Hill
operator obtained by Dubrovin, Its-Matveev and
McKean-van Moerbeke.
\end{abstract}

\section {Introduction and main results}
\setcounter{equation}{0}

In the present paper we determine the trace formula for the fourth
order operator on the circle. Before we recall the famous trace
formula for the second order operator \cite{D}, \cite{IM},
\cite{MM}. Consider the differential equation
\[
\lb{E1}
-y''+qy=\l y,\qqq \l  \in \C,
\]
where $q$ is the 1-periodic potential and $\C$ is a complex plane.
Denote by $\a_{0}^+<\a_{1}^-\le\a_{1}^+<\a_{2}^-\le...$ eigenvalues
of the equation \er{E1} with the 2-periodic boundary conditions.

Consider the equation \er{E1} with the Dirichlet  boundary conditions for
the shifted potential $q(\cdot+t)$ by any fix parameter $t\in\T=\R/\Z$.
Denote the corresponding eigenvalues by  $\b_1(t)<\b_2(t)<...$. It is well-known that
\[
\lb{ldper}
\b_n(t )\in [\a_n^-,\a_n^+]\qqq\forall\qq(n,t )\in\N\ts\T.
\]
Now we can recall the  famous  trace formula, i.e., the identity given by
\[
\lb{trf}
q(t)=\a_0^++\sum_{n=1}^\iy\Big(\a_{n}^++\a_{n}^-
-2\b_n(t)\Big)\qqq\forall\qq t \in\T.
\]
Dubrovin \cite{D} and Its~--~Matveev \cite{IM} determined the  trace
formula \er{trf} for the so-called $N$ band  potential, when
$\b_n(t)=\a_n^-=\a_n^+$ for all $n\ge N,  t\in\T$ and some $N\ge 1$.
In this case the sum in \er{trf} is finite. McKean~--~van Moerbeke
\cite{MM} and Trubowitz \cite{T} determined the  trace formula
\er{trf} for sufficiently smooth potential. Korotyaev \cite{K}
determined the trace formula \er{trf} for the case $q\in L^2(\T)$.

We discuss the trace formula for the fourth order operator.
Introduce the Sobolev space $W_m^s(A)$, where $A$ is the finite
interval or the circle,  by
\[
W_m^s(A)=\{f,  f^{(m)}\in L^s(A)\},\qqq s\ge 1,\qq m=0,1,2,..
\]

 Our  main goal is to extend the  trace formula \er{trf} for
the fourth order equation
\[
\lb{E2}
y''''+2(py')'+qy=\l y,\qqq   \qq \l \in \C,
\]
where the 1-periodic real functions $p,q$ satisfy the conditions
\[
\lb{Cpq} p\in W_4^{1}(\T),\qqq q\in  W_2^{1}(\T).
\]

Consider the equation \er{E2} with the 2-periodic  boundary
conditions. Let $\l_0^+,\l_{n}^\pm,n\in\N$, be the corresponding
spectrum labeled by $ \l_0^+\le\l_1^-\le\l_1^+\le\l_2^-\le..., $
counted with multiplicities. Note that the eigenvalues have
multiplicity $\le 4$. The following asymptotics hold true:
\[
\lb{4g.T2-2i}
\l_n^\pm=(\pi n)^4-2 p_0(\pi n)^2+{ p_0^2-\|p\|^2\/2}+q_0
+o(n^{-{3\/2}}),
\]
as $n\to+\iy$, see \cite{BK1}, where
$$
f_0=\int_0^1 f(t)dt,\qq
\|f\|^2=\int_0^1|f(t)|^2dt.
$$

Consider the equation \er{E2} with the shifted coefficients  by any
fix parameter $t\in\R$:
\[
\lb{E2t}
y''''+2(p(x+t)y')'+q(x+t)y=\l y,\qqq   \qq \l \in \C,
\]
with the Dirichlet type boundary conditions
\[
\lb{E2tbc}
 y(0)=y''(0)=y(1)=y''(1)=0.
\]
 Let $\m_n(t),n\ge 1$, be the corresponding
spectrum labeled by $ \m_1(t)\le\m_2(t)\le\m_3(t)\le..., $ counted
with multiplicities. Note that the eigenvalues have multiplicity
$\le 2$. The following asymptotics hold true:
\[
\lb{4g.asDir}
\m_n(t)=(\pi n)^4-2p_0(\pi n)^2+{p_0^2-\|p\|^2\/2}+q_0+{O(1)\/n^{2}}
\]
as $n\to+\iy$ uniformly on $t\in\T$, see \cite{BK2}.

Now we present trace formulas, which are similar to the second order
case \er{trf}.

\begin{theorem}
\lb{4g.thasdir} Let $(p,q)\in W_4^1(\T)\ts W_2^1(\T)$. Then the
following  identity
\[
\lb{4g.trf3}
q(x)-{p''(x)\/2}=
\l_{0}^++\sum_{n=1}^\iy\big(\l_{n}^++\l_{n}^--2\m_{n}(x)\big)\qqq\forall\qq x \in  \T,
\]
holds true, where the series converges absolutely and uniformly on $x\in \T$.

In particular, if $p=\const$, then
\[
\lb{4g.trf30}
q(x)=\l_{0}^++\sum_{n=1}^\iy\big(\l_{n}^++\l_{n}^--2\m_{n}(x)\big)\qqq\forall\qq x \in\T,
\]
where the series converges absolutely and uniformly on $x\in \T$.
\end{theorem}

There are numerous results about the trace formulas for the higher
order operators, see McKean \cite{McK}, Ostensson
\cite{O}, Sadovnichii \cite{S1}, \cite{S2},
Sadovnichii and Podol'skii
\cite{SP}, Zatitskiy, Nazarov and Stolyarov \cite{ZNS} and  see
references therein. The inverse spectral results for the fourth
order operators on a finite interval were obtained by Caudill, Perry
and Schueller \cite{CPS}, McLaughlin \cite{McL},  Papanicolaou and
Kravvaritis \cite{PK}. The spectral problem for the fourth and
higher order operator with periodic coefficients considered by
Badanin and Korotyaev \cite{BK3}, \cite{BK4}, Mikhailets and
Molyboga \cite{MMo}, Papanicolaou \cite{P95}, \cite{P03}, Tkachenko
\cite{Tk}.

\section {The resolvents}
\setcounter{equation}{0}

Let $\cB$, $\cB_1$, $\cB_2$ be the set of all bounded, trace class
and Hilbert-Schmidt class operators on $L^2(\R)$,
respectively.
The norms of  $\cB$, $\cB_1$, $\cB_2$ are denoted by $\|\cdot\|, \
\|\cdot\|_1, \ \|\cdot\|_2$, respectively.

\subsection{The second order operators.}
We recall the following well known results about the second order
operators $h_1,h_2$, see, for instance, in \cite{MW}, \cite{LS}.

Firstly, we consider the operator $h_2y=-y''-py$ acting in $L^2(0,2)$
with the
2-periodic boundary conditions.
The spectrum of $h_2$ is discrete and its eigenvalues
$\a_0^+,\a_{n}^\pm,n\in\N$, are simple or have multiplicity two.
They can enumerated, counting with multiplicity, so that
$\a_0^+<\a_{1}^-\le\a_{1}^+<\a_{2}^-\le\a_{2}^+<...$.
The following asymptotics  hold true:
\[
\lb{asevH}
\a_n^\pm=(\pi n)^2+O(1)
\qqq\as\qq n\to+\iy.
\]

Secondly, we consider the operator $h_1$ acting in $L^2(0,1)$ with
the Dirichlet  boundary conditions given by
\[
\lb{4g.domp}
h_1y=-y''-py,\qqq  y(0)=y(1)=0.
\]
 All eigenvalues $\b_n,n\in\N$, of the operator
$h_1$ are simple. We enumerate their in the increasing order
$\b_1<\b_2<...$ Recall that $ \b_n\in[\a_n^-,\a_n^+]$ for all
$n\ge1$. Moreover, the following trace formula holds true:
\[
\lb{S01}
p^2(0)+{p''(0)\/2}=(\a_{0}^+)^2+\sum_{n\ge1}\big((\a_{n}^+)^2+(\a_{n}^-)^2-2\b_{n}^2\big)
\]
where the series converges absolutely, see \cite[p.254]{MM}.

\subsection{The fourth order operators}
 Introduce the self-adjoint operator $H_2$  in $L^2(0,2)$ by
\[
\begin{aligned}
H_2=\pa^4+2\pa p\pa +q \qq  {\rm on} \qq [0,2],
\end{aligned}
\]
with 2-periodic boundary conditions.

It is sufficiently to prove identity \er{4g.trf3} at $x=0$. Thus we
need to define the operator $H_1$ in $L^2(0,1)$ by
$$
\begin{aligned}
H_1=\pa^4+2\pa p\pa +q\qq {\rm on} \  [0,1],
\end{aligned}
$$
with the Dirichlet type boundary conditions
$$
f(0)=f''(0)=f(1)=f''(1)=0.
$$

Using the identity
\[
\lb{V}
\pa^4+2\pa p\pa +q=(-\pa^2-p)^2+V,\qqq V=q-p''-p^2,
\]
and $V\in L^\iy(0,1)$,  we obtain
\[
\lb{Hjhj}
H_j=h_j^2+V,\qqq j=1,2,
\]
where the operators $h_1$
and $h_2$ are introduced above.
Let the operator $h_j^0, j=1,2$, be equal
to the operator $h_j$ at $p=0$. Introduce the resolvents
\[
\lb{resolv} R_j(\l)=(H_j-\l)^{-1},\qq \cR_j(\l)=(h_j^2-\l)^{-1},\qq
\cR_j^0(\l)=\big((h_j^0)^2-\l\big)^{-1},
\]
where $j=1,2.$ We have
\[
\lb{cRj}
\cR_j=\cR_j^0-\cR_j(-h_j^0p-ph_j^0+p^2)\cR_j^0.
\]
Due to asymptotics \er{4g.T2-2i}, \er{4g.asDir}, \er{asevH}  all
resolvents satisfy
\[
\cR_j(\l), \ R_j(\l),\  h_j^0\cR_j^0(\l) \in \cB_1,
\]
on the corresponding resolvent sets. Define the contours $K_n\ss \C$
by
\[
\textstyle K_n=\big\{\l\in \C:|\l|^{1\/4}=\pi
\big(n+{1\/2}\big)\big\}, \qqq n\ge 1.
\]
We present results about the asymptotics, proved in Section 3.

\begin{lemma}
\lb{TL1} Let $j=1,2$ and let $n\to \iy$. Then the following
asymptotics hold true:
\[
\lb{res11}
\|\cR_j^0(\l)\|_2+\|\cR_j(\l)\|_2+
\|R_j(\l)\|_2=O(n^{-3}),
\]
\[
\lb{res12} \|h_j^0\cR_j^0(\l)\|_2=O(n^{-1})
\]
uniformly  on the contours $K_n$, and
\[
\lb{TrVR}
\oint_{K_n}\Big(\l\Tr V\cR_j^2(\l)+\Tr V\cR_j^0(\l)\Big)d\l=o(1).
\]

\end{lemma}

\subsection {Proof of the main results}
Introduce the function
\[
\lb{Phi} \F=\Tr( R_2-\cR_2)-2\Tr( R_1-\cR_1).
\]

\begin{lemma} The following identity holds true:
\[
\lb{iFV} \lim_{n\to+\iy} {1\/2\pi i}  \oint_{K_n}\l \F(\l)d\l
=-V(0).
\]
\end{lemma}

\no {\bf Proof.} Substituting the identities
$$
R_j=\cR_j- R_jV\cR_j =\cR_j-\cR_jV\cR_j+R_jV\cR_jV\cR_j,\qq j=1,2,
$$
into \er{Phi} we obtain
\[
\lb{Fi} \F=\F_0+\F_1,
\]
where
\[
\lb{Fi01} \F_0=-\Tr V\cR_2^2+ \Tr 2V\cR_1^2, \qqq \F_1=\Tr
V\cR_2V\cR_2R_2- \Tr 2V\cR_1V\cR_1R_1.
\]

Let $n\to\iy$. Identities \er{Fi01}, \er{TrVR} yield
$$
\oint_{K_n}\big(\l\F_0(\l)-F(\l)\big)d\l=o(1),
$$
where
\[
\lb{F}
F=\Tr V\cR_2^0-2\Tr V\cR_1^0.
\]
Estimates \er{res11} give
$$
|\Tr V\cR_j(\l)V\cR_j(\l)R_j(\l)|\le
\|R_j(\l)\|_2\|\cR_j(\l)\|_2^2\|V\|^2=O(n^{-9}),\qq j=1,2,
$$
uniformly on contours $K_n$, which yields $ \F_1(\l)=O(n^{-9}) $ and
then
$$
 \oint_{K_n} \l\F_1(\l)d\l=o(1).
$$
Then identity \er{Fi} gives
\[
\lb{FFi} \oint_{K_n}\big(\l\F(\l)-F(\l)\big)d\l=o(1).
\]

We have the Fourier series
\[
\lb{Fs}
V(x)=V_0+2\sum_{n=1}^\iy \Big(V_{cn}\cos 2\pi nx+ V_{sn}\sin
2\pi nx\Big),
\]
where
$$
V_{sn}=\int_0^1V(x)\sin 2\pi nxdx, \qq V_{cn}=\int_0^1V(x)\cos2\pi
nxdx, \qq V_0=\int_0^1V(x)dx.
$$

Let $e_n={1\/\sqrt 2}e^{i\pi nx}, n\in \Z$, and $s_n=\sin\pi nx,
n\in\N$. Define the scalar products $\lan f,g\ran=\int_0^2f\ol
gdx$ and $(f,g)=\int_0^1f\ol gdx$ in $L^2(0,2)$ and $L^2(0,1)$
respectively. Identity \er{F} gives
$$
\begin{aligned}
F(\l)={\lan Ve_0,e_0\ran \/-\l}+\sum_{n=1}^\iy {\lan V e_n,e_n\ran
+\lan Ve_{-n},e_{-n}\ran -4(Vs_n,s_n) \/(\pi
n)^4-\l}=-{V_0\/\l}+\sum_{n=1}^\iy{2V_{cn}\/(\pi n)^4-\l},
\end{aligned}
$$
since we have the identities
$$
\lan V e_n,e_n\ran=V_0\qqq\forall\qq n\in\Z,
$$
$$
(Vs_n,s_n)=\int_0^1V\sin^2\pi nxdx ={V_0-V_{cn}\/2},\qq n\ge 1.
$$
Then
$$
{1\/2\pi i}\oint_{K_N}F(\l)d\l=-V_0-2\sum_{n=1}^N V_{cn} \qqq
$$
for all $N\ge 1$. Identity \er{FFi} give
$$
{1\/2\pi i}\oint_{K_N}\l\F(\l)d\l=-\Big(V_0+2\sum_{n=1}^N
V_{cn}\Big)+o(1)\qqq\as\qq N\to\iy,
$$
then \er{Fs} yields \er{iFV}. $\BBox$

\no {\bf Proof of Theorem \ref{4g.thasdir}.} Asymptotics
\er{4g.T2-2i}, \er{4g.asDir} show that the series
$$
S=\l_{0}^++\sum_{n\ge1}\big(\l_{n}^++\l_{n}^--2\m_{n}\big)
$$
converges absolutely. Let $S_0$ be given by
$$
S_0=(\a_{0}^+)^2+\sum_{n\ge1}\big((\a_{n}^+)^2+(\a_{n}^-)^2-2\b_{n}^2\big).
$$
Then due to \er{Phi} we have
$$
S-S_0 =-{1\/2\pi i}\lim_{n\to+\iy}\oint_{K_n} \l\F(\l)d\l.
$$
Identity \er{S01} yields
\[
\lb{im1} S-p^2(0)-{p''(0)\/2} =-{1\/2\pi
i}\lim_{n\to+\iy}\oint_{K_n} \l\F(\l)d\l.
\]
Substituting identity \er{iFV} into \er{im1} we obtain
$$
S-p^2(0)-{p''(0)\/2} =V(0).
$$
Substituting \er{V} into the last identity we have
$S=q(0)-{p''(0)\/2}$, which yields \er{4g.trf3}.
 $\BBox$

\section {Proof of Lemma \ref{TL1}}
\setcounter{equation}{0}

 \no {\bf Proof of Lemma \ref{TL1}.}
 We will prove \er{res11} for $R_1(\l)$. The
proof for other $R_2, \cR_j, \cR_j^0$ is similar. Let
$$a_k=|\m_k|^{1\/4},\qq  a=|\l|^{1\/4},\qq \l\in K_n,\ \qq k,n\ge 1.
$$
Then we have
\[
\lb{estm1}
|\m_k-\l|\ge \big||\m_k|-|\l|\big|
=|a_k-a|(a_k+a)(a_k^2+a^2)\ge |a_k-a|a^3.
\]
Asymptotics \er{4g.asDir} yields
\[
\lb{}
 a_k= \pi k+ \pi\ve_k,\qqq \where \qq |\ve_k|<{1\/4}\qq \forall \ k>
 N,
\]
\[
\lb{estm2} |a_k-a|\ge 1\qqq \forall\qq k\le N
\]
for $n\in\N$ large enough and
 for some $N\in\N$ large enough. We have
\[
\lb{estm3}
|a_k-a|=\pi\Big|k-n-{1\/2}+\ve_k\Big|\ge\pi\Big(|k-n|-{1\/4}\Big)\qqq\forall\qq k> N.
\]
Estimates \er{estm1} -- \er{estm3} give
$$
\|R_1(\l)\|_2^2=\sum_{k=1}^\iy{1\/|\m_k-\l|^2}\le
{1\/a^6}\sum_{k=1}^\iy{1\/|a_k-a|^2}\le{1\/a^6}
\Big(N+\sum_{k=N+1}^{\iy}{1\/\pi^2(|k-n|-{1\/4})^2}\Big)\le
 {C\/a^6},
$$
where $C=N+{1\/\pi^2}\sum_{k\in\Z}(k-{1\/4})^{-2}<\iy$ and this
yields \er{res11} for $R_1(\l)$.

We will prove \er{res12} for $j=1$. The proof for $j=2$ is similar.
We have
\[
\lb{est1}
|(\pi k)^4-\l|\ge\big||(\pi k)^4-|\l|\big|
=|\pi k-a|(\pi k+a)\big((\pi k)^2+a^2\big)\ge a|\pi k-a|\big((\pi k)^2+a^2\big).
\]
This estimate implies
$$
\|h_j^0\cR_j^0(\l)\|_2^2=\sum_{k=1}^\iy{(\pi k)^4\/|(\pi k)^4-\l|^2}
\le{1\/a^2}\sum_{k=1}^\iy{1\/|\pi k-a|^2}
\le{1\/\pi^2 a^2}\sum_{k=1}^\iy{1\/|k-n-{1\/2}|^2}
\le{C_1\/a^2},
$$
where $C_1={1\/\pi^2 }\sum_{k\in \Z}(k-{1\/2})^{-2}<\iy$,
which yields \er{res12} for $j=1$.

\medskip

We will prove asymptotics \er{TrVR} for $j=2$. The proof for $j=1$
is similar. The integration by parts gives
$$
\oint_{K_n}\l\Tr V\cR_2^2(\l)d\l =\oint_{K_n}\l\Tr V(\cR_2(\l))'d\l
=-\oint_{K_n}\Tr V\cR_2(\l)d\l
$$
for all $n\in\N$. Using identity \er{cRj} we obtain
\[
\lb{T1}
\begin{aligned}
\oint_{K_n}\Big(\l\Tr V\cR_2^2(\l)+\Tr V\cR_2^0(\l)\Big)d\l
=\oint_{K_n}\Tr V\big(\cR_2^0(\l)-\cR_2(\l)\big)d\l
\\
=\oint_{K_n}\Tr \big(V\cR_2(\l)Q\cR_2^0(\l)\big)d\l, \qqq\where\qq
Q=-h_2^0p-ph_2^0+p^2.
\end{aligned}
\]
Asymptotics \er{res11}, \er{res12} give
\[
\lb{T13}
\big|\Tr\big(\cR_2^0(\l)Q(\l)\cR_2^0(\l)Q(\l)\cR_2(\l)\big)\big|
\le\|\cR_2^0(\l)Q(\l)\|_2^2\|\cR_2(\l)\|_2
=O(n^{-5})\qq\as\qq n\to\iy
\]
uniformly on all contours $K_n$.
Substituting
\er{cRj} into \er{T1} and using \er{T13} we obtain
\[
\lb{T11} \oint_{K_n}\Big(\l\Tr V\cR_2^2(\l)+\Tr V\cR_2^0(\l)\Big)d\l
=\oint_{K_n}\Tr V\cR_2^0(\l)Q\cR_2^0(\l)d\l+o(1)\qq\as\qq n\to\iy.
\]

Thus we need to consider $\Tr V\cR_2^0(\l)Q\cR_2^0(\l)$, where
$Q=p^2-h_2^0p-ph_2^0$. Firstly, we consider the case $p^2$. Estimate
\er{res11} as $n\to\iy$ yields
\[
\lb{T14} \oint\limits_{K_n} \Tr V\cR_2^0(\l)p^2\cR_2^0(\l)d\l= O(1)
\oint\limits_{K_n}\|\cR_2^0(\l)\|_2^2 |d\l| =\oint\limits_{K_n}
O(n^{-6}) |d\l|=O(n^{-2}).
\]

Secondly, we consider  the case  $-h_2^0p-ph_2^0$. Using $ \wh
f_k=\int_0^2f(x)e^{i\pi kx}dx $ and  the identity
$$
\lan (-h_2^0p-ph_2^0)e_m,e_k \ran= \int_0^2e^{-i\pi
kx}(\pa_x^2p(x)+p(x)\pa_x^2)e^{i\pi mx}dx =-\pi^2(k^2+m^2)\wh
p_{m-k},
$$
where
$e_m={1\/\sqrt2}e^{i\pi mx}$ and $\lan f,g\ran=\int_0^2f\ol gdx$,
we have
\[
\lb{r3}
\begin{aligned}
\Tr(V\cR_2^0(h_2^0p+ph_2^0)\cR_2^0) =-\sum_{m,k=-\iy}^\iy F(k,m,\l)
\\
=-\sum_{k=-\iy}^\iy {8\pi^2V_0p_0k^2\/((\pi k)^4-\l)^2}
-\sum_{m,k=-\iy\atop m\ne k}^\iy F(k,m,\l),\qq
F(k,m,\l)={\pi^2(k^2+m^2)\wh {V}_{m-k}\wh {p}_{m-k} \/((\pi
m)^4-\l)((\pi k)^4-\l)},
\end{aligned}
\]
where the series converge uniformly on each contour $K_n,n\in\N$.
Moreover, the identity $\oint_{K_n} { d\l\/((\pi k)^4-\l)^2}=0$ and the
decomposition $\{|k|\ne |m|\}=D_1\cup D_2\cup D_3$, where  $D_1,
D_2,  D_3$ are given by
$$
\begin{aligned}
D_1=\{|k|\ne |m|, |k|\le n, |m|\le n\}\cup \{|k|\ne |m|, |k|> n, |m|> n\},\\
D_2=\{|k|\le n, |m|> n\},\qqq \qq     D_3=   \{|k|> n, |m|\le n\},\\
\end{aligned}
$$
give
\[
\lb{KnI}
\oint_{K_n}\Tr(V\cR_2^0(h_2^0p+ph_2^0)\cR_2^0) d\l= \oint_{K_n}
\sum_{m,k=-\iy\atop m\ne k}^\iy F(k,m,\l)d\l=I_1(\l)+I_2(\l)+I_3(\l)
\]
for all $n\in\N$,
where
$$
I_j(\l)=\oint_{K_n} \sum_{D_j} F(k,m,\l)d\l.
$$
We have $I_1(\l)=0$ and thus we need to consider
$I_2, I_3$.

Consider $I_3$, the proof for $I_2$ is similar. Identity \er{KnI}
gives
\[
\lb{T15}
I_3(\l)=\oint_{K_n} \sum_{D_3} F(k,m,\l)d\l= {2 \/i\pi}
\sum_{D_3}{\wh {V}_{m-k}\wh {p}_{m-k} \/m^2-k^2}.
\]
Consider the case $k>n,|m|\le n$, the proof for the other case is
similar. Using $V\in W_2^1(0,2)$ and $p\in W_4^1(0,2)$  we obtain
\[
{|\wh {V}_{m-k}\wh {p}_{m-k}| \/| m^2-k^2|}\le
{C\/|k-m|^6 | m^2-k^2|}\le {C\/|k-m|^{7} (| m|+|k|)}
\]
for some constant $C$. Define $k'=k-n\ge 1 $ and $ m'=n-m\in
[0,2n]$. Then we obtain
$$
 {1\/|k-m|^{7} (| m|+|k|)}\le {1\/|k'+m'|^{7} n}.
$$
This yields
$$
\sum_{k> n, |m|\le n}{|\wh {V}_{m-k}\wh {p}_{m-k}|\/|m^2-k^2|} \le
\sum_{k'\ge 1, m'\ge 0} {C\/|k'+m'|^{7} n}={C_1\/n},\qq C_1=
\sum_{k'\ge 1, m'\ge 0} {C\/|k'+m'|^{7}}.
$$
Similar arguments show that
$$
\sum_{k<-n, |m|\le n}{|\wh {V}_{m-k}\wh {p}_{m-k}|\/|m^2-k^2|}
\le{C_2\/n}
$$
for some $C_2>0$ and then \er{T15} yields $I_3(\l)=O(n^{-1})$ as
$n\to\iy.$  Similar estimates  yield $I_2(\l)=O(n^{-1})$.
Then \er{KnI} gives
$$
 \oint_{K_n}\Tr(V\cR_2^0(h_2^0p+ph_2^0)\cR_2^0) d\l=O(n^{-1}).
$$
Substituting this asymptotics and
\er{T14} into \er{T11} we obtain \er{TrVR}.
$\BBox$

\

 \setlength{\itemsep}{-\parskip} \footnotesize
 \no
{\bf Acknowledgments.}
{Various parts of this paper were written
during Evgeny Korotyaev's stay in Mittag-Leffler Institute, Sweden
and Centre for Quantum Geometry of Moduli spaces (QGM),  Aarhus
University, Denmark. He is grateful to the institutes for the
hospitality. His study was supported by the Ministry of education
and science of Russian Federation, project 07.09.2012  No 8501 and
the RFFI grant "Spectral and asymptotic methods for studying of the
differential operators" No 11-01-00458 and partly supported by the
Danish National Research Foundation grant DNRF95 (Centre for Quantum
Geometry of Moduli Spaces - QGM)"}

\end{document}